# Licensing Open Government Data

*Jyh-An Lee*[*]

Governments around the world create and collect an enormous amount of data that covers important environmental, educational, geographical, meteorological, scientific, demographic, transport, tourism, health insurance, crime, occupational safety, product safety, and many other types of information.[1] This data is generated as part of a government's daily functions.[2] Given government data's exceptional social and economic value, former U.S. President Barack Obama described it as a "national asset."[3] For various policy reasons, open government data ("OGD") has become a popular governmental practice and international movement in recent years.[4] Open government has therefore acquired a new meaning empowered by digital technologies and data science.[5]

---

\* Assistant Professor at the Faculty of Law in the Chinese University of Hong Kong; J.S.D., Stanford Law School; LL.M., Harvard Law School. I would like to thank Ann S. Chiu, Tyng-Ruey Chuang, Brett Frischmann, Richard Gruner, Mike Madison, Sean Pager, Chung-Lun Shen, Ming-Yan Shieh, Victoria Stodden, Haochen Sun, Qian Wang, Zhang Ping, and participants in the "Museum Computing: An Approach to Bridging Cultures, Communities and Science" conference in National Palace Museum, Taipei, "Taking Data into the Public Domain" conference at the University of Macau, "Chengdu International Copyright Fortune Summit" held by the Chengdu Municipal Government, 2016 Intellectual Property Scholar Conference ("IPSC") at Stanford Law School, 2016 Annual Cross-Strait Four-Region Forum on Copyright in Taipei held by Fujen School of Law and Peking University Law School and Intellectual Property School, and Asia-Pacific Copyright Association ("APCA")'s 2016 conference on "Copyright in the Asian Pacific: The Challenges and Opportunities" at the University of Hong Kong Faculty of Law for helpful comments on earlier drafts. I am grateful to Lesley Luo Jingyu for her research assistance.

1. Keiran Hardy & Alana Maurushat, *Opening Up Government Data for Big Data Analysis and Public Benefit*, 33 COMPUTER L. & SECURITY REV. 30, 31 (2017); NAT'L ARCHIVES, UK GOVERNMENT LICENSING FRAMEWORK FOR PUBLIC SECTOR INFORMATION 5 (5.0 ed. 2016), *available at* http://www.nationalarchives.gov.uk/documents/information-management/uk-government-licensing-framework.pdf [hereinafter NAT'L ARCHIVES, UK GOVERNMENT LICENSING FRAMEWORK]; JOSHUA TAUBERER, THE PRINCIPLES AND PRACTICES OF OPEN GOVERNMENT DATA 10 (2nd ed. 2014); Katleen Janssen, *The Influence of the PSI Directive on Open Government Data: An Overview of Recent Developments*, 28 GOV'T INFO. Q. 446, 446 (2011); Barbara Ubaldi, *Open Government Data: Towards Empirical Analysis of Open Government Data Initiatives* 4 (Organisation for Economic Cooperation and Development, Working Papers on Public Governance No. 22, 2013), *available at* http://www.oecd-ilibrary.org/governance/open-government-data_5k46bj4f03s7-en.

2. Miriam Marcowitz-Bitton, *Commercializing Public Sector Information*, 97 J. PAT. & TRADEMARK OFF. SOC'Y 412, 413 (2015).

3. The White House Office of the Press Secretary, *Obama Administration Releases Historic Open Data Rules to Enhance Government Efficiency and Fuel Economic Growth*, OBAMA WHITE HOUSE ARCHIVE (May 9, 2013), *available at* https://www.whitehouse.gov/the-press-office/2013/05/09/obama-administration-releases-historic-open-data-rules-enhance-government.

Numerous national and local governments in Australia, Brazil, Canada, Ireland, New Zealand, Taiwan, the United Kingdom (UK), and the United States (US), etc., have implemented new policies to release their data or encourage people to gain access to, use, and reuse, government data.[6] It is estimated that more than 250 national or local governments from around 50 developed and developing countries have launched OGD initiatives.[7] Data.gov, established by the United States federal government, and Data.gov.uk, launched by the British government, are both notable examples of data portals through which governments make their data available to the public.[8]

In July 2013, G8 leaders signed the G8 Open Data Charter, which outlined five fundamental open data principles.[9] Two years earlier, the international OGD movement had led to the establishment of the Open Data Partnership (ODP), "a multilateral initiative that aims to secure concrete commitments from governments to promote transparency, empower citizens, fight corruption, and harness new technologies to strengthen governance."[10] The ODP was initiated by eight national governments (Brazil, Indonesia, Mexico, Norway, the Philippines, South Africa, the UK, and the US) with the proclamation of the Open Government Declaration on September 20, 2011.[11] Sixty-two additional national governments have joined the ODP since its incorporation.[12] Moreover, seventy governments altogether have made more than 2,250 commitments to implement open data policies.[13] International organizations, such as the World Bank, have also actively advocated for and implemented open data policies.[14]

---

4. *See, e.g.*, JOEL GURIN, OPEN DATA NOW: THE SECRET TO HOT STARTUPS, SMART INVESTING, SAVVY MARKETING, AND FAST INNOVATION 216–18 (2014); Anneke Zuiderwijk & Marijn Janssen, *Open Data Policies, Their Implementation and Impact: A Framework for Comparison*, 31 GOV'T INFO. Q. 17, 17 (2014); Teresa Scassa, *Public Transit Data Through an Intellectual Property Lens: Lessons About Open Data*, 41 FORDHAM URB. L.J. 1759, 1760–61 (2014); DEIRDRE LEE ET AL., OPEN DATA IRELAND: BEST PRACTICE HANDBOOK 26 (2014), per.gov.ie/wp-content/uploads/Best-Practice-Handbook.pdf.

5. *See, e.g.*, Jillian Raines, Note, *The Digital Accountability and the Transparency Act of 2011 (DATA): Using Open Data Principles to Revamp Spending Transparency Legislation*, 57 N.Y.L. SCH. L. REV. 313, 321–24 (2013); Nataša Veljković et al., *Benchmarking Open Government: An Open Data Perspective*, 31 GOV'T INFO. Q. 278, 278–80 (2014); *see also* Jeremy Weinstein & Joshua Goldstein, *The Benefits of A Big Tent: Opening Up Government in Developing Countries: A Response to Yu & Robin's The New Ambiguity of "Open Government"*, 60 UCLA L. REV. DISCOURSE 38, 40–41 (2012) (noting the distinction between and the convergence of the "technologies for open data" and "politics of open government.").

6. *See, e.g.*, GURIN, *supra* note 4, at 217–19; LEE ET AL., *supra* note 4, at 37–44, 228–29; TAUBERER, *supra* note 1, at 17–18, 24, 89-91; Hardy & Maurushat, *supra* note 1, at 33-34; Harlan Yu & David G. Robinson, *The New Ambiguity of "Open Government"*, 59 UCLA L. REV. DISCOURSE 178, 198–200 (2012); Tiago Peixoto, *The Uncertain Relationship Between Open Data and Accountability: A Response to Yu and Robinson's the New Ambiguity of "Open Government"*, 60 UCLA L. REV. DISCOURSE 200, 210 (2013); KENT MEWHORT, CREATIVE COMMONS LICENSES: OPTIONS FOR CANADIAN OPEN DATA PROVIDERS, 5–7 (June 1, 2012), *available at* cippic.ca/sites/default/files/Creative%20Commons%20Licenses%20-%20Options%20for%20Canadian%20Open%20Data%20Providers.pdf; *An Outline of the Government Open Data Promotion Situation in Taiwan*, NATIONAL DEVELOPMENT COUNCIL, *available at* http://www.ndc.gov.tw/en/News_Content.aspx?n=8C362E80B990A55C&sms=1DB6C6A8871CA043&s=09819AC7E099BCD5 (last visited Mar. 15, 2017).

7. *Open Data in 60 Seconds*, THE WORLD BANK, *available at* opendatatoolkit.worldbank.org/en/open-data-in-60-seconds.html (last visited Mar. 15, 2017).

8. GURIN, *supra* note 4, at 10–11, 218; Yu & Robinson, *supra* note 6, at 198, 200; *see also* JEAN-LOUIS MONINO & SORAYA SEDKAOUI, BIG DATA, OPEN DATA AND DATA DEVELOPMENT xxxv (2016) (noting that these are the two leading nations globally in promoting open data policies); Esteve Sanz, *Open Governments and Their Cultural Transitions*, *in* OPEN GOVERNMENT: OPPORTUNITIES AND CHALLENGES FOR PUBLIC GOVERNANCE 1, 11 (Mila Gascó-Hernández ed., 2014) (describing the role of Data.gov).

9. *FAQ*, OPEN DATA CHARTER, *available at* opendatacharter.net/faq/ (last visited Mar. 15, 2017).

Businesses are also embracing the open data trend as reflected in new strategies, applications, products, and services. For example, Microsoft introduced the "Open Government Data Initiative" to promote the company's Window Azure online platform as a tool for OGD.[15] Government data has become an increasingly important strategic source for entrepreneurship, innovation, and economic growth.[16] Businesses may aggregate, repack, and redistribute the data, develop new applications and platforms, combine the data with other information, or explore novel ways to add value to government data. Enterprises can make use of such data to provide services relating to travel, business planning, shopping advice, etc.[17] The commercial value of this volume of government data is increasingly apparent in the Big Data technology environment.[18] A number of nonprofit organizations (NPOs), such as the Open Data Institute, Open Knowledge Foundation, and the Sunlight Foundation, have also actively taken part in the OGD movement in different ways.[19]

---

10. *What Is the Open Government Partnership?*, OPEN GOVERNMENT PARTNERSHIP, *available at* opengovpartnership.org/about (last visited Mar. 15, 2017).
11. *Id.*
12. *Id.*
13. *Id.*
14. *See, e.g.*, *World Bank Open Data*, WORLD BANK, *available at* data.worldbank.org/ (last visited Mar. 15, 2017).
15 *See, e.g.*, Steve Clayton, *Microsoft's Open Government Data Initiative with Windows Azure*, MICROSOFT DEVELOPER (May 11, 2009), *available at* blogs.msdn.microsoft.com/stevecla01/2009/05/11/microsofts-open-government-data-initiative-with-windows-azure/; Marius Oiaga, *Windows Azure Powers Microsoft Open Government Data Initiative*, SOFTPEDIA (May 7, 2009), *available at* news.softpedia.com/news/Windows-Azure-Powers-Microsoft-Open-Government-Data-Initiative-111061.shtml.
16. *See infra* Section II.B.
17. *See, e.g.*, Frederik Zuiderveen Borgesius et al., *Open Data, Privacy, and Fair Information Principles: Towards a Balancing Framework*, 30 BERKELEY TECH. L.J. 2073, 2081 (2015).
18. *See, e.g.*, MONINO & SEDKAOUI, *supra* note 8, at 30–33, 38; Michael Chui et al., *Generating Economic Value Through Open Data*, *in* BEYOND TRANSPARENCY: OPEN DATA AND THE FUTURE OF CIVIC INNOVATION 163, 163 (Brett Goldstein & Lauren Dyson eds., 2013); Joel Gurin, *Big Data and Open Data: How Open Will the Future Be?*, 10 I/S: J.L. & POL'Y FOR INFO. SOC'Y 691, 699–700 (2015); Hardy & Maurushat, *supra* note 1, at 30-31; Ubaldi, *supra* note 1, at 5–7; *see also* Maureen K. Ohlhausen, *The Social Impact of Open Data*, FEDERAL TRADE COMMISSION (July 23, 2014), *available*

210         *HASTINGS BUSINESS LAW JOURNAL*         [Vol. 13:2

OGD policy involves various legal issues, ranging from personal data protection,[20] citizens' right of access to government information or freedom of information,[21] the attribution of legal liability,[22] and appropriate parties to release government data.[23] Intellectual property (IP) licensing has both been viewed as a cornerstone for OGD,[24] and, from a cynical perspective, as one of the main obstacles to the release of governments' open data.[25] Entrepreneurs may hesitate to use or reuse government data if there is no reliable licensing or clear legal arrangement governing it.[26] Tim Berners-Lee, inventor of the Internet, provided a Five-Star Scheme to evaluate the degree of dataset reusability.[27] The scheme's initial One-Star level sets the most fundamental requirement for OGD, which is that data should be accessible online under an open license.[28] However, this scheme neither illustrates what is an appropriate open license for OGD nor explains why an open license matters for OGD.

This Article focuses on legal issues associated with OGD licenses. Different government agencies with different policy goals have set different licensing terms to release their data.[29] These licensing terms

---

*at* ftc.gov/system/files/documents/public_statements/571281/140723socialimpactofopendata.pdf (addressing the relationship between Big Data and OGD from the perspective of U.S. Federal Trade Commission).

   19.   *See, e.g.*, LEE ET AL., *supra* note 4, at 28–31.

   20.   *See, e.g.*, GURIN, *supra* note 4, at 183–95, 232; Micah Altman et al., *Towards a Modern Approach to Privacy-Aware Government Data Releases*, 30 BERKELEY TECH. L.J. 1967, 2005, 2048–59 (2015); Hardy & Maurushat, *supra* note 1, at 34; Jeff Jonas & Jim Harper, *Open Government: The Privacy Imperative*, *in* OPEN GOVERNMENT: COLLABORATION, TRANSPARENCY, AND PARTICIPATION IN PRACTICE 315 (Daniel Lathrop & Laurel Ruma eds., 2010); Borgesius et al., *supra* note 17, at 2086–93, 2107–14, 2125–29; Mashael Khayyat & Frank Bannister, *Open Data Licensing: More Than Meets the Eye*, 20 INFO. POL'Y 231, 244–45 (2015); Zuiderwijk & Janssen, *supra* note 4, at 22, 26; LEE ET AL., *supra* note 4, at 63–65; Ubaldi, *supra* note 1, at 43.

   21.   *See, e.g.*, TAUBERER, *supra* note 1, at 87–89, 125; Jeffrey D. Rubenstein, *Hacking FOIA: Requests to Drive Government Innovation*, *in* BEYOND TRANSPARENCY: OPEN DATA AND THE FUTURE OF CIVIL INNOVATION 81 (Brett Goldstein & Lauren Dyson eds., 2013); Gurin, *supra* note 18, at 700–01; Marcowitz-Bitton, *supra* note 2, at 419–23; LEE ET AL., *supra* note 4, at 21; Ubaldi, *supra* note 1, at 4–5, 37.

   22.   Zuiderwijk & Janssen, *supra* note 4, at 22.

   23.   *See generally* David Robinson et al., *Government Data and the Invisible Hand*, 11 YALE J.L. & TECH. 160 (2009) (arguing that the private sector, commercial or nonprofit organizations, rather than the government, is better suited to deliver OGD).

   24.   *See, e.g.*, Ubaldi, *supra* note 1, at 37.

   25.   *See, e.g.*, Khayyat & Bannister, *supra* note 20, at 232; *see also* Janssen, *supra* note 1, at 452 (noting that quite a few French governments had been struggling with licensing policies toward OGD).

   26.   Ubaldi, *supra* note 1, at 11.

   27.   Michael Hausenblas & James J. Kim, 5 STAR OPEN DATA, *available at* 5stardata.info/en/ (last visited Mar. 15, 2017).

   28.   *Id.*

   29.   *See, e.g.*, GOVERNMENT REFORM UNIT, DEPARTMENT OF PUBLIC EXPENDITURE AND REFORM IN IRELAND, OPEN DATA LICENCES 9 (2015), *available at* per.gov.ie/wp-content/uplo

reflect policy considerations that differ from those contemplated in business transactions or shared in typical commons scenarios, such as free or open source software communities.[30] They also concern some fundamental IP issues that are not covered by, or analyzed in depth in, the current literature. The aim of this Article is to provide a comprehensive legal analysis of open data licenses. This study argues that the choice and design of an open data license forms an important element of a government's information policy. Part I introduces the concept and characteristics of OGD, which emphasizes citizens' easy and timely access to government data. The features associated with OGD have begun to form an increasingly universal principle adhered to around the world. Part II identifies the primary policy goals of OGD, which include the enhancement of governmental transparency, accountability, public participation, the improvement of democracy and public service quality, and the advancement of innovation and economic development. These policy goals should be the deciding factors in the design and choice of license. Part III explores the most prevalent or notable standardized open data licenses adopted by governments worldwide. These licenses were drafted by Creative Commons, Open Data Commons, and the British government. A brief analysis of the terms of these licenses is also provided therein. Part IV examines the major legal issues pertaining to the licensing of OGD. As a large portion of government data is factual information automatically generated by machines or software, it fails to meet the originality standard and thus cannot be protected by copyright. Consequentially, the licensing of such public domain data becomes legally contentious, especially in jurisdictions that do not provide *sui generis* protection of databases. Moreover, Part IV discusses the attribution provision, which is the most common restriction in OGD licenses. Based on theories of moral right, Part IV also explains the rationale behind attribution provisions and non-endorsement provisions in open data licenses. Part V concludes.

## I. The Concept of Open Data

OGD, sometimes referred to as open public sector information (PSI),[31] represents policies or practices that make data held by the public sector digitally available and accessible for reuse or redistribution for free or at a

---

ads/Open-Data-Licence-Consultation-Paper-February-2015.docx; Zuiderwijk & Janssen, *supra* note 4, at 26; Mewhort, *supra* note 6, at 2–3.

    30. *See, e.g.*, Jyh-An Lee, *New Perspectives on Public Goods Production: Policy Implications of Open Source Software*, 9 Vand. J. Ent. & Tech. L. 45, 50–53 (2009); Molly Shaffer Van Houweling, *The New Servitudes*, 96 Geo. L.J. 885, 925–26 (2008).

    31. Zuiderwijk & Janssen, *supra* note 4, at 17.

nominal cost.[32] According to the European Union (EU) Directive on the Re-Use of Public Sector Information, "[o]pen data policies . . . encourage the wide availability and re-use of public sector information for private or commercial purposes, with minimal or no legal, technical, or financial constraints."[33] Commentators may link the open data movement to other similar movements in which information is liberalized and widely disseminated by digital technologies and the Internet.[34] Those movements include open access, open educational resources, open standard, and free/open source software initiatives.[35]

    A number of organizations and individuals have provided their own definitions of, or criteria for, open data. For example, a working group led by Carl Malamud first attempted to set eight principles for open data in December 2007; these principles include: (1) complete; (2) primary; (3) timely; (4) accessible; (5) machine processable; (6) non-discriminatory; (7) non-proprietary; and (8) license-free.[36] Open Knowledge International (OKI), a nonprofit network advocating for free access to, and the sharing of, information globally, defines open data as "data that can be freely used, re-used and redistributed by anyone—subject only, at most, to the requirement to attribute and [sic] sharealike."[37] According to OKI's definition, there should be no discrimination against the different uses of government data.[38] Therefore, "'non-commercial' restrictions that would prevent 'commercial' use, or restrictions of use for certain purposes (e.g., only in education), are not allowed."[39] Moreover, the Sunlight Foundation

---

    32. *See, e.g.*, Hardy & Maurushat, *supra* note 1, at 30; TAUBERER, *supra* note 1, at 95; GOVERNMENT REFORM UNIT, DEPARTMENT OF PUBLIC EXPENDITURE AND REFORM IN IRELAND, *supra* note 29, at 4; Ubaldi, *supra* note 1, at 6; *see also* GURIN, *supra* note 4, at 9 ("Open Data can best be described as accessible public data that people, companies, and organizations can use to launch new ventures, analyze patterns and trends, make data-driven decisions, and solve complex problems"); Borgesius et al., *supra* note 17, at 2075 ("[o]pen government data refers to data released by public sector bodies, in a manner that is legally and technically re-usable"). *But see* Luca Leone, *Open Data and Food Law in the Digital Era: Empowering Citizens Through ICT Technology*, 10 EUR. FOOD & FEED L. REV. 356, 358 (2015) (claiming that there is no generally accepted definition of open data or OGD).
    33. Directive 2013/37/EU of the European Parliament and of the Council of 26 June 2013, 2013 O.J. (L 175) 1, 1 Amending Directive 2003/98/EC on the Re-Use of Public Sector Information, 2003 O.J. (L 345), 90 [hereinafter EU Amending Directive on Re-Use of PSI].
    34. *See, e.g.*, Yu & Robinson, *supra* note 6, at 187–88; *see also* Sanz, *supra* note 8, at 3–5, 8–11 (describing a series of openness movements enabled by the internet, including the free/open source software movement, and their relations to OGD).
    35. *See, e.g.*, TAUBERER, *supra* note 1, at 12–13, 93; Scassa, *supra* note 4, at 1779–80; Yu & Robinson, *supra* note 6, at 187–88; *see also* Weinstein & Goldstein, *supra* note 5, at 40 (noting that OGD represents "a new alignment of open source and transparency").
    36. Joshua Tauberer, *The Annotated 8 Principles of Open Government Data*, OPENGOVDATA.ORG (Dec. 8, 2007) *available at* opengovdata.org/; TAUBERER, *supra* note 1, at 187–88; Khayyat & Bannister, *supra* note 20, at 342; Ubaldi, *supra* note 1, at 8.
    37. Open Knowledge International, *What is Open Data*, OPEN DATA HANDBOOK, *available at* open datahandbook.org/guide/en/what-is-open-data/ (last visited Mar. 15, 2017).
    38. *Id.*
    *39. Id.*

also published its *Open Data Policy Guidelines* to illustrate OGD best practices.[40] The International Open Data Charter, which is based on the G8 Open Data Charter, identifies six principles of open data: (1) open by default; (2) timely and comprehensive; (3) accessible and usable; (4) comparable and interoperable; (5) for improved governance and citizen engagement; and (6) for inclusive development and innovation.[41] In sum, other than being timely, comprehensive, and openly accessible,[42] below are two other noteworthy principles supported by most open data advocates:

OPEN BY DEFAULT

Many believe that it is a general principle that government data should be openly and freely available online, whereas the non-disclosure of government data should be an exception. A government's proactive disclosure of data is essential to its transparency and democratic governance.[43] In other words, governments shall open their data by default unless there is a compelling reason, such as national security or privacy protection, to keep the data confidential.[44] This principle is recognized in the G8 Open Data Charter[45] and in the open data policies of the European Union and the UK.[46] It can also be found in New York City's *Technical Standards Manual*, which states that "[a]ll public data sets must be considered open unless they contain information designated as sensitive, private, or confidential as defined by the Citywide Data Classification Policy or information that is exempt pursuant to the Public Officers Law, or any other provision of a federal or state law, rule or regulation or local law."[47]

OPEN FORMAT

Government data should be made available in formats for all types of use.[48] The data should be in formats that are machine readable, downloadable, usable, and distributable.[49] Such formats are typically open or non-proprietary industrial protocols and formats.[50] Put differently, "[a]n open format is one that is platform independent, machine readable, and made available to the public without restrictions that would impede the re-use of that information."[51] Extensible Markup Language (XML) is an example of open format enabling interoperability of data from diverse sources.[52]

---

40. *Open Data Policy Guidelines*, SUNLIGHT FOUNDATION, *available at* sunlightfoundation.com/opendataguidelines/#proactive-release (last visited Mar. 15, 2017).
41. *Principles*, OPEN DATA CHARTER, *available at* opendatacharter.net/principles/ (last visited Mar. 15, 2017) [hereinafter, Open Data Charter, *Principles*].
42. *See, e.g.*, TAUBERER, *supra* note 1, at 98–99, 115; Ubaldi, *supra* note 1, at 8–9, 24.
43. Borgesius et al., *supra* note 17, at 2084.
44. *See, e.g.*, GURIN, *supra* note 4, at 219.
45. Open Data Charter, *Principles*, *supra* note 41.
46. *See, e.g.*, GURIN, *supra* note 4, at 219.
47. New York City Department of Information Technology & Telecommunications, *City Policies*, NYC OPENDATA TECHNICAL STANDARDS MANUAL (2012), *available at* cityofnewyork.github.io/opendatatsm/citypolicies.html (last visited Mar. 15, 2017).
48. TAUBERER, *supra* note 1, at 99.
49. *See, e.g.*, Raines, *supra* note 5, at 324; Ubaldi, *supra* note 1, at 24.

In Tim Berners-Lee's Five-Star Open Data Scheme, "using non-proprietary formats" is at the Three-Star level.[53] An open format can effectively promote the analysis and reuse of government data.[54] The Obama Administration endorsed releasing government data in "computer-readable" forms.[55] Similarly, the Open Government Declaration, the United States and seven other signatory countries committed in September 2011 to "provide high-value information, including raw data, in a timely manner, in formats that the public can easily locate, understand and use, and in formats that facilitate reuse."[56]

## II. Policy Goals Underlying Open Data

OGD brings important social, economic, and democratic value to society.[57] Likewise, it can promote both public and private interests.[58] An EU Directive on the Re-Use of Public Sector Information further highlights the value of open data policies:

> Open data policies . . . which promote the circulation of information not only for economic operators but also for the public, can play an important role in kick-starting the development of new services based on novel ways to combine and make use of such information, stimulate economic growth and promote social engagement.[59]

---

50. TAUBERER, *supra* note 1, at 99; *see also* Teresa Scassa & Robert J. Currie, *New First Principles? Assessing the Internet's Challenges to Jurisdiction*, 42 GEO. J. INT'L L. 1017, 1067 (noting that efforts to control format in the OGD settings are fading).
51. PETER R. ORSZAG, EXEC. OFFICE OF THE PRESIDENT, MEMORANDUM NO. M-10-06, OPEN GOVERNMENT DIRECTIVE 2 (2009), *available at* treasury.gov/open/Documents/m10-06.pdf.
52. *See, e.g.*, Leone, *supra* note 32, at 358.
53. Hausenblas & Kim, *supra* note 27.
54. TAUBERER, *supra* note 1, 100–01.
55. *Technology*, OBAMA WHITE HOUSE ARCHIVE, *available at* obamawhitehouse.archives.gov/issues/ technology (last visited Mar. 15, 2017).
56. *Open Government Declaration*, OPEN GOV'T P'SHIP (Sept. 2011), *available at* opengovpartnership.org/siteswww.opengovpartnership.org/files/page_files/OGP_Declaration.pdf.
57. *See, e.g.*, Pater Conradie & Sunli Choenni, *On the Barrier for Local Government Releasing Open Data*, 31 GOV'T INFO. Q. 10, 10 (2014).
58. *See, e.g.*, GURIN, *supra* note 4, at 218.
59. EU Amending Directive on Re-Use of PSI, *supra* note 33.

Open data is tasked with changing the way people run governments and do business via freely available government data.[60] Therefore, the aim of OGD policies is to build an ecosystem with multiple functions. Identifying policy goals for OGD and setting priorities are also critically important for the design of data governance and relevant legal structures, including licenses. Here in part two, we briefly analyze the policy goals underlying OGD.

*Transparency and Accountability*

OGD promotes the transparency of government and the policymaking process, which underpins accountability and democracy.[61] Transparency involves the disclosure of actions taken by the public sector.[62] Government data can definitely shed light on government activities. Some government data, such as that pertaining to public spending, distribution of revenue, and subsidy, is critically important for government accountability.[63] Therefore, by enabling the monitoring of government activities, open data can help reduce corruption.[64] The Obama Administration has identified its open data policy goal as increasing transparency, participation, and collaboration,[65] which will eventually advance the quality and efficiency of the services provided by the government.[66] Likewise, the French government's OGD policy aims to promote government accountability and make good use of the "collective intelligence of its citizens."[67] The Australian government similarly acknowledged how public access and the reuse of government information could enhance public participation and 4216  HASTINGS BUSINESS LAW JOURNAL  [Vol. 13:2democracy.[68] The World Bank stated that open data "encourages greater citizen participation in government affairs" and "supports democratic societies."[69] Therefore, OGD has been viewed as a tool to advance public scrutiny, political accountability,[70] participation, and the quality of government services. In return, all of these benefits will improve the governance of, and trust in, the public sector.[71]

---

60. GURIN, *supra* note 4, at 9.
61. *See, e.g.*, GOVERNMENT REFORM UNIT, DEPARTMENT OF PUBLIC EXPENDITURE AND REFORM IN IRELAND, *supra* note 29, at 4; Keiran Hardy & Alana Maurushat, *supra* note 1, at 33; TAUBERER, *supra* note 1, at 132; Chui et al., *supra* note 18, at 163–64; Janssen, *supra* note 1, at 446; Leone, *supra* note 32, at 356, 358; Marcowitz-Bitton, *supra* note 2, at 413; Scassa, *supra* note 4, at 1760; Veljković et al., *supra* note 5, at 280; Yu & Robinson, *supra* note 6, at 196–97; Zuiderwijk & Janssen, *supra* note 4, at 17; Ubaldi, supra note 1, at 4, 11–12; Weinstein & Goldstein, *supra* note 5, at 46; *see also* Judith Bannister, *Open Government: From Crown Copyright to the Creative Commons and Culture Change*, 34 U. N. S. WALES L.J. 1080, 1089 (stating that open access to government information improves transparent decision making and the quality of democracy); Peixoto, *supra* note 6, at 202, 207 (arguing that open date enables transparency, which may lead to accountability).
62. Peixoto, *supra* note 6, at 203.
63. Borgesius et al., *supra* note 17, at 2083.
64. *See, e.g.*, Marcowitz-Bitton, *supra* note 2, at 416; *Starting an Open Data Initiative*, WORLD BANK, *available at* opendatatoolkit.worldbank.org/en/starting.html (last visited Mar. 15, 2017).
65. Borgesius et al., at 2083–84; Sanz, *supra* note 8, at 10.
66. Keiran Hardy & Alana Maurushat, *supra* note 1, at 32; ORSZAG, *supra* note 51, at 1; Scassa, *supra* note 4, at 1760; Yu & Robinson, *supra* note 6, at 196, 201; Zuiderwijk & Janssen, *supra* note 4, at 17; *see also* Borgesius et al., *supra* note 17, at 2085–86 (illustrating how open data promotes public sector efficiency and improves the quality of public service); Peixoto, *supra* note 6, at 202 (arguing that OGD enables participation, which fosters better services and policies).
67. Borgesius et al., *supra* note 17, at 2083.

*Economic Development*

OGD has been viewed as a crucial strategy to build a "data-driven economy."[72] The immense volume and diversity of government data may bring great commercial value to enterprises.[73] Put more clearly, OGD is an abundant free resource that fuels a wide range of new innovative products, apps, services, and business models associated with data reuse and analysis.[74] Additional value is then created "by means of crowdsourcing, user tracking, and data analytics."[75] Various commercial uses of government data may further encourage economic development.[76] In other words, a properly designed OGD policy can unlock the value of PSI to the public sector.[77] On May 9, 2013, when the US Office of Management and Budget and the Office of Science and Technology Policy announced the Open Data Policy, President Obama signed an Executive Order to promote OGD and stated that

> [Open data can] fuel more private sector innovation . . . . And talented entrepreneurs are doing some pretty amazing things with it . . . . Starting today, we're making even more

---

68. Bannister, *supra* note 61, 1091–92.
69. World Bank, *supra* note 64.
70. Yu & Robinson, *supra* note 6, at 182.
71. Open Data Charter, *Principles*, *supra* note 41.
72. Leone, *supra* note 32, at 358.
73. *See, e.g.*, Leone, *supra* note 32, at 356; Marcowitz-Bitton, *supra* note 2, at 413; Scassa, *supra* note 4, at 1773–74 (describing the commercial value of transit data); Gurin, *supra* note 18, at 693–96.
74. *See, e.g.*, GURIN, *supra* note 4, at 23–35, 218–19; Chui et al., *supra* note 18, at 163, 168; Janssen, *supra* note 1, at 446; Marcowitz-Bitton, *supra* note 2, at 416; *see also* Conradie & Choenni, *supra* note 57, at 10 (stating that "the release of [government] data for a broader use may give a boost to the creative industry, which in return leads to innovative applications and techniques").
75. Michael Halberstam, *Beyond Transparency: Rethinking Election Reform from an Open Government Perspective*, 38 SEATTLE U. L. REV. 1007, 1028 (2015).
76. *See, e.g.*, GOVERNMENT REFORM UNIT, DEPARTMENT OF PUBLIC EXPENDITURE AND REFORM IN IRELAND, *supra* note 29, at 4; Scassa, *supra* note 4, at 1760–61; Zuiderwijk & Janssen, *supra* note 4, at 17; *see also* GURIN, *supra* note 4, at 217 (stating that open government datasets "can have a powerful impact for the public good and economic growth"); Gianluca Misuraca, et al., *Policy-Making 2.0: Unleashing the Power of Big Data for Public Governance*, in OPEN GOVERNMENT: OPPORTUNITIES AND CHALLENGES FOR PUBLIC GOVERNANCE 171, 171 (Mila Gascó-Hernández ed., 2014) (describing the benefit brought by OGD in the commercial field).
77. *See, e.g.*, Borgesius et al., *supra* note 17, at 2080.

> government data available online, which will help launch even more new startups. And we're making it easier for people to find the data and use it, so that entrepreneurs can build products and services we haven't even imagined yet.[78]

The European Commission (EC) also highlighted the potential for significant economic gains to come from OGD.[79] Similarly, both the UK[80] and Australian[81] governments have stated that OGD could greatly benefit the economy. A number of studies have estimated that the economic value brought by OGD will exceed hundreds of millions, or even trillions, of dollars.[82] For example, the McKinsey Global Institute estimated that open data can unlock an economic value of $3–5 trillion a year across seven sectors in the US.[83] In summary, OGD can form an important part of a government's economic policy when it comes to fostering innovation and economic development.

### III. Standardized Licenses for OGD

Some OGD advocates believe that true open data should be free from license restrictions;[84] others claim that without specific open licenses, it is too costly for users to search and negotiate with data publishers.[85] For those who believe licenses are necessary for OGD, the consensus is that the licenses, or terms and conditions, should facilitate optimal access to the underlying data.[86] Government agencies may choose click-use or standardized licenses, such as a Creative Commons (CC) license,[87] or develop their own licensing terms.[88] The primary advantage of using standardized licenses is to save costs associated with creating a bespoke license, to achieve order and efficiency, and to achieve interoperability between licenses.[89] In this section, we introduce the most common public licenses considered by governments for OGD and analyze their similarities and dissimilarities.

---

78. The White House Office of the Press Secretary, *supra* note 3.
79. Zuiderwijk & Janssen, *supra* note 4, at 17.
80. *See, e.g.*, GURIN, *supra* note 4, at 9; NAT'L ARCHIVES, UK GOVERNMENT LICENSING FRAMEWORK, *supra* note 1, at 6; Janssen, *supra* note 1, at 451.
81. Bannister, *supra* note 61, at 1091.
82. *See, e.g.*, Borgesius et al., *supra* note 17, at 2082; LEE ET AL., *supra* note 4, at 18–19; Chui et al., *supra* note 18, at 166; Marcowitz-Bitton, *supra* note 2, at 424; Ubaldi, *supra* note 1, at 15.
83. McKinsey Global Institute, *Open Data: Unlocking Innovation and Performance with Liquid Information*, MCKINSEY & COMPANY (Oct. 2013), *available at* mckinsey.com/business-functions/business-technology/our-insights/open-data-unlocking-innovation-and-performance-with-liquid-information.
84. *See, e.g.*, TAUBERER, *supra* note 1, at 106, 144–45; Yu & Robinson, *supra* note 6, at 196; *see also* Yochai Benkler, *Book Review: Commons and Growth: The Essential Role of Open Commons in Market Economies*, 80 U. CHI. L. REV. 1499, 1551 (2013) (claiming that OGD is subject to no constraint).
85. *See, e.g.*, GOVERNMENT REFORM UNIT, DEPARTMENT OF PUBLIC EXPENDITURE AND REFORM IN IRELAND, *supra* note 29, at 6; *see also* Federico Morando, *Legal Interoperability: Making Open Government Data Compatible with Businesses and Communities*, 4 ITALIAN J. LIBR. ARCHIVES & INFO. SCI. 441, 442 (2013) (introducing the viewpoint that "the distribution of data also requires . . . licensing").
86. Ruth Okediji, *Government as Owners of Intellectual Property? Considerations for Public Welfare in the Era of Big Data*, 18 VAND. J. ENT. & TECH. L. 331, 336 (2016).
87. *See, e.g.*, Marcowitz-Bitton, *supra* note 2, at 439; LEE ET AL., *supra* note 4, at 67.

*Creative Commons*

Creative Commons (CC) is a nonprofit organization that enables users to donate their works to the public domain or to freely license their works under certain conditions.[90] CC provides a suite of standardized copyright licenses and has been playing an important role in the global movement advocating for information sharing and reuse.[91] CC licenses have always been an option for OGD policies.[92] In the "Guidelines on Recommended Standard Licences, Datasets and Charging for the Reuse of Documents," the European Commission recommended CC BY and CC0 for OGD:

> Open standard licences, for example the most recent Creative Commons (CC) licences (version 4.0), could allow the re-use of PSI without the need to develop and update custom-made licences at the national or sub-national level. Of these, the CC0 public domain dedication is of particular interest.[93]

In this section, we will analyze the options provided by CC for OGD.

---

88. *See, e.g.*, Food and Drug Administration (FDA), *Terms of Services*, OPEN FDA (Mar. 22, 2014), *available at* open.fda.gov/terms/ (last visited Mar. 15, 2017).

89. *See, e.g.*, Khayyat & Bannister, *supra* note 20, at 238; Marcowitz-Bitton, *supra* note 2, at 434; NAOMI KORN & CHARLES OPPENHEIM, LICENSING OPEN DATA: A PRACTICAL GUIDE 4 (ver. 2.0 June 2011), *available at* discovery.ac.uk/files/pdf/Licensing_Open_Data_A_Practical_Guide.pdf; *see also* Mewhort, *supra* note 6, at 2, 9–10 (noting the benefit of interoperability brought by CC licenses).

90. *See, e.g.*, LAWRENCE LESSIG, FREE CULTURE: HOW BIG MEDIA USES TECHNOLOGY AND THE LAW TO LOCK DOWN CULTURE AND CONTROL CREATIVITY 283–84 (2003); ROBERT P. MERGES, JUSTIFYING INTELLECTUAL PROPERTY 86 (2011); Timothy K. Armstrong, *Shrinking the Commons: Termination of Copyright Licenses and Transfers for the Benefit of the Public*, 47 HARV. J. ON LEGIS. 359, 382 (201); Jyh-An Lee, *The Greenpeace of Cultural Environmentalism*, 16 WIDENER L. REV. 1, 12–13 (2010).

91. *See, e.g.*, JAMES BOYLE, THE PUBLIC DOMAIN: ENCLOSING THE COMMONS OF THE MIND 182–83 (2008); LAWRENCE LESSIG, REMIX: MAKING ART AND COMMERCE THRIVE IN THE HYBRID ECONOMY 276–79 (2008); Armstrong, *supra* note 90, at 383–84; Clark D. Asay, *A Case for the Public Domain*, 74 OHIO ST. L.J. 753, 759 (2013); Lee, *supra* note 90, at 13.

92. *See, e.g.*, TAUBERER, *supra* note 1, at 89 (describing that San Francisco's open data law requiring "generic license" such as a CC license).

93. Commission Notice, Guidelines on Recommended Standard Licences, Datasets and Charging for the Reuse of Documents, 2014 O.J. (C 240) 1, 1, § 2.2 [hereinafter Commission Notice, Guidelines on Recommended Standard Licences].

*Creative Commons Licenses*

CC licenses provide copyright owners with the option of making creative works available for reproduction, distribution, and other use by granting some exceptions to copyright law.[94] CC licenses used to cover only copyrighted works,[95] but started to cover database rights after version 4.0 was released in 2013. CC licenses consist of one or more of the following four main elements: (1) Attribution (BY) requires the licensee to give credit to the licensor; (2) NonCommercial (NC) prohibits licensees from using the work for commercial purposes; (3) NoDerivatives (ND) forbids the licensee from adapting the work; and (4) ShareAlike (SA) requires licensees to license the adapted works under the same CC licenses.[96] Based on these four elements, CC offers six combination licenses: (1) CC-BY; (2) CC-BY-SA; (3) CC-BY-NC; (4) CC-BY-ND; (5) CC-BY-NC-SA; and (6) CC-BY-NC-ND.[97]

CC-BY, the least restrictive CC license, is currently the most popular standardized license among EU Member States.[98] The Australian government also uses CC-BY to release its data as default.[99] Users may use the data for commercial or non-commercial purpose as long as they provide attribution.[100] Some government agencies in Australia adopt more restrictive CC license that prevents users from making derivative works and/or commercial uses.[101] The New Zealand Government Open Access and Licensing framework (NZGOAL) likewise recommended CC-BY license for government works.[102]

---

CC0

CC0 is not a license *per se*.[103] Rather, CC0 is a statement surrendering copyright and related rights, such as database rights, worldwide and permanently.[104] If the waiver is not legally effective for any reason, CC0 acts as a license, granting everyone an unconditional, irrevocable, nonexclusive, royalty free license to use the work.[105] Some open data advocates believe that CC0 is the best tool to release government data.[106] The government of the Netherlands has launched a website which uses CC0 to waive the copyright on government data.[107] The Norwegian government also adopted CC0 to open its data.[108] Although CC0 is not a common practice for OGD, it occasionally attracts the support of OGD advocates.[109]

PDM

A Public Domain Mark (PDM) is a mark provided by CC.[110] PDM's can be used to mark and tag content that is already in the public domain and is not subject to any copyright restriction.[111] The nature of PDM is more akin to a declaration than a contract. PDM is different from CC0. The former can be used by anyone for works that are in the public domain; whereas the latter is intended to be used by copyright holders or other related right holders for their underlying proprietary works.[112] In addition, CC0 transforms proprietary content into content in the public domain, whereas PDM does not change the legal status of materials that are already in the public domain.[113]

---

103. *See. e.g.*, TAUBERER, *supra* note 1, at 110. *But see* Christopher S. Brown, Comment, *Copyleft, The Disguised Copyright: Why Legislative Copyright Reform Is Superior to Copyleft Licenses*, 78 UMKC L. REV. 749, 774 (2010) (suggesting that the mechanism of CC0 is "more of a license" than PDM introduced in subsection II.A.3 below); Emily Hudson & Robert Burrell, *Abandonment, Copyright and Orphaned Works: What Does It Mean to Take the Proprietary Nature of Intellectual Property Rights Seriously?*, 35 MELB. U. L. REV. 971, 996 (2011) (viewing CC0 as a license); Khayyat & Bannister, *supra* note 20, at 240 (defining CC0 as a license).
104. *CC0 Use for Data*, CREATIVE COMMONS, *available at* wiki.creativecommons.org/wiki/CC0_use_for_data (last visited Mar. 15, 2017).
105. *CC0 1.0 Universal*, CREATIVE COMMONS, *available at* creativecommons.org/publicdomain/zero/1.0/legalcode (last visited Mar. 15, 2017).
106. *See, e.g.*, TAUBERER, *supra* note 1, at 110; Janssen, *supra* note 1, at 451.
107. Mike, *New Dutch Government Portal Uses CC0 Public Domain Waiver as Default Copyright Status*, CREATIVE COMMONS (Mar. 31, 2010), *available at* creativecommons.org/2010/03/31/new-dutch-government-portal-uses-cc0-public-domain-waiver-as-default-copyright-status/; *Case Studies/ Netherlands Government*, CREATIVE COMMONS, *available at* wiki.creativecommons.org/wiki/Case_Studies/Netherlands_Government (last visited Mar. 15, 2017).
108. Mewhort, *supra* note 6, at 16.
109. *See, e.g.*, Janssen, *supra* note 1, at 454; Mewhort, *supra* note 6, at 2; Joshua Tauberer et al., *Open Government Data: Best-Practices Language for Making Data "License-Free"*, @UNITEDSTATES PROJECT (Dec. 12, 2013), *available at* theunitedstates.io/licensing/.
110. *Public Domain Mark 1.0*, CREATIVE COMMONS, *available at* creativecommons.org/publicdomain/mark/1.0/ (last visited Mar. 15, 2017).
111. *Public Domain Mark*, CREATIVE COMMONS, *available at* creativecommons.org/share-your-work/public-domain/pdm/ (last visited Mar. 15, 2017).

OPEN DATA COMMONS

The Open Data Commons (ODC) project, under the Open Knowledge Foundation, has been sponsored by an information management company, Talis, to provide legal tools for sharing data.[114] The project was initiated by Jordan Hatcher and Professor Charlotte Waelde in 2007 with the aim of providing standardized licenses for a *sui generis* database right in EU countries because CC 3.0 did not cover database rights at that time.[115] ODC created three solutions specifically for data, datasets, and databases:

1. Open Data Commons Attribution License (ODC-BY)

An ODC-BY license covers both database right and copyright.[116] Similar to CC-BY, the only requirement of the licensee is that the licensee shall attribute any public use of the database, or works produced from the database, in the manner specified in the license.[117]

2. Open Data Commons Open Database License (ODbL)

Like all ODC licenses, ODbL includes database right and copyright.[118] However, ODbL imposes more restrictions on licensees. Like CC-BY-SA, ODbL has a share-alike requirement that obliges the licensee to adopt the same license for the adapted database or adapted works created by the licensee.[119] Additionally, if the licensee redistributes the database or creates an adapted version with technological measures on it, he or she must redistribute a version without such measures in place.[120]

---

112. *PDM FAQ*, CREATIVE COMMONS, *available at* wiki.creativecommons.org/wiki/PDM_FAQ (last visited Mar. 15, 2017).

113. *Id.*

114. *See, e.g.*, Lucie Guibault & Thomas Margoni, *Analysis of Licensing Issues*, in SAFE TO BE OPEN: STUDY ON THE PROTECTION OF RESEARCH DATA AND RECOMMENDATIONS FOR ACCESS AND USAGE 154 (Lucie Guibault & Andreas Wiebe eds., 2013); Guadamuz & Cabell, *supra* note 95, at 17; Morando, *supra* note 85, at 444–45; Khayyat & Bannister, *supra* note 20, at 238; Jordan Hatcher, *Implementing Open Data: The Open Data Commons Project*, TECH. INNOVATION & MGMT. REV. (Feb. 2008), http://timreview.ca/article/122.

115. Marinos Papadopoulos & Charalampos Bratsas, *Openness/Open Access for Public Sector Information and Works—The Creative Commons Licensing Model* 12 (European Public Sector Information Platform, Topic Report No. 2015/06, June 2015), *available at* europeandataportal.eu/sites/default/files/2015_open_access_for_public_sector_information_and_works.pdf.

116. *Open Data Commons Attribution License (ODC-By) v1.0*, OPEN DATA COMMONS, *available at* opendatacommons.org/licenses/by/1.0/. (last visited Mar. 15, 2017) [hereinafter Open Data Commons, *ODC-By v1.0*].

117. *Id.*

118. *Open Database License (ODbL) v1.0*, OPEN DATA COMMONS, *available at* opendatacommons.org/licenses/odbl/1.0/ (last visited Mar. 15, 2017) [hereinafter Open Data Commons, *ODbL*].

119. *Id.*

### 3. Public Domain Dedication License (PDDL)

PDDL is a standardized waiver of copyright and database right.[121] PDDL is CC0's counterpart in the ODC project.[122] Similar to CC0, if PDDL is not legal (for any reason), it serves as a public license without any restrictions.[123] PDDL has been adopted by the City of Surrey in Canada to release its data.[124]

## UK GOVERNMENT LICENSING FRAMEWORK FOR PUBLIC SECTOR INFORMATION (UKGLF)

Some governments design the public licenses for their own government data. For example, France has developed the License Ouverte,[125] Germany created Datenlizenz Deutschland,[126] and Italy adopted the Italian Open Data License.[127] The British government has developed the UK Government Licensing Framework for Public Sector Information (UKGLF) to license "the use and re-use of public sector information both in central government and the wider public sector."[128] The UKGLF provides three licensing schemes for government data, which might be the most notable OGD licenses drafted by the government:

1. The Open Government Licence (OGL), which permits "free use and re-use for all purposes, both commercial and non-commercial".[129]

2. The Non-Commercial Government Licence, which applies to "free use and re-use for non-commercial purposes only".[130] This restriction is similar to the licensing arrangement of CC-BY-NC.

---

120. *Id.*
121. *ODC Public Domain Dedication and License (PDDL)*, OPEN DATA COMMONS, *available at* opendatacommons.org/licenses/pddl/1.0/ (last visited Mar. 15, 2017) [hereinafter Open Data Commons, *PDDL*].
122. For information regarding CC0 see text accompanying note 104; Guibault & Margoni, *supra* note 114, at 155; Mewhort, *supra* note 6, at 16.
123. Open Data Commons, PDDL, *supra* note 121.
124. Mewhort, *supra* note 6, at 2, 16.
125. *See, e.g.*, Khayyat & Bannister, *supra* note 20, at 241; *Licence Ouverte Open Licence*, ETALAB, *available at* etalab.gouv.fr/wp-content/uploads/2014/05/Open_Licence.pdf (last visited Mar. 15, 2017); *Open Licence (French)*, WIKIPEDIA, *available at* en.wikipedia.org/wiki/Open_licence_(French) (last visited Mar. 15, 2017).
126. *See, e.g.*, LEE ET AL., *supra* note 4, at 69.
127. *See, e.g.*, Janssen, *supra* note 1, at 452; Mockus & Palmirani, *supra* note 145, at 293–94.
128. NAT'L ARCHIVES, UK GOVERNMENT LICENSING FRAMEWORK, *supra* note 1, at 4.
129. *Id.*
130. *Id.*

3. The Charged Licence, "where charges are made for the re-use of information."[131] Different from other public licenses for OGD, the Charged Licence allows for payment to be made by users or licensees. However, governments can only charge for overhead for producing the document.[132]

OGL, which is similar to a CC-BY or ODC-BY license, is the default license of UKGLF.[133] In other words, a general principle in the United Kingdom is that government data should be freely available for both commercial and noncommercial purposes as long as attribution is given. As a result, OGL has been embraced by quite a few ministerial and local government agencies.[134] Different from CC and ODC-BY licenses, OGL addresses "Crown copyright,"[135] which is tailored for the UK copyright regime.[136]

*Analysis of the Terms*

The open licenses introduced above share a number of similarities. Excepting CC0, PDM, and PDDL, all other licenses require attribution. With the exception of a Charged Licence, licensees under other licenses can freely use the licensed materials. CC-BY, ODC-BY, and OGL are the least restrictive licenses in which the licensors still retain their IP rights. ODbL has a counterpart in CC licenses, which is the CC-BY-SA license—it has a similar share-alike requirement for licensees.[137] Nonetheless, there are still significant differences between these two licenses. On the one hand, a CC-BY-SA license simply restricts licensees from applying technological measures,[138] whereas ODbL provides a more flexible option for dual licensing.[139] On the other hand, ODbL only covers the database

---

131. *Id.*

132. The Re-Use of Public Sector Information Regulations 2015, SI 2015/1415, § 15(2) (UK), *available at* legislation.gov.uk/uksi/2015/1415/contents/made.

133. NAT'L ARCHIVES, UK GOVERNMENT LICENSING FRAMEWORK, *supra* note 1, at 5, 15.

134. Okediji, *supra* note 86, at 351.

135. Some countries other than the United Kingdom, such as Australia and Canada, also have the Crown copyright rules governing copyright over government information. *See, e.g.*, Elizabeth F. Judge, *Copyright, Access, and Integrity of Public Information*, 1 J. PARLIAMENTARY & POL. L. 427, 427–28 (2008); Ann L. Monotti, *Nature and Basis of Crown Copyright in Official Publications*, 14(9) EUR. INTELL. PROP. REV. 305, 305 (1992).

136. *Open Government Licence for Public Sector Information*, NAT'L ARCHIVES, *available at* nationalarchives.gov.uk/doc/open-government-licence/version/3/ (last visited Mar. 15, 2017). [Hereinafter Nat'l Archives, *OGL Version 3*].

137. *Attribution-ShareAlike 4.0 International*, CREATIVE COMMONS, *available at* creativecommons.org/ licenses/by-sa/4.0/legalcode (last visited Mar. 15, 2017).

138. *Id.*

139. *See supra* text accompanying note 120; Open Data Commons, ODbL, *supra* note 118, § 4.7 ("a. This License does not allow You to impose (except subject to Section 4.7 b.) any terms or any technological measures on the Database, a Derivative Database, or the whole or a Substantial part of the Contents that alter or restrict the terms of this License, or any rights granted under it, or have the effect itself, and not its content.[140] Put differently, ODbL only covers the manner in which the data is selected or arranged and the database as a whole; ODbL does not cover each individual item of data or content. If the licensee would also like to license individual pieces of content, he or she needs to apply for another license instrument. OpenStreetMap, an open source map project contributed to by volunteers,[141] provides a good example to demonstrate the distinction between ODbL and CC-BY-SA. The project licenses its whole database to the public under ODbL.[142] But when it comes to the licensing arrangement of its individual copyrighted map, OpenStreetMap needs to implement a CC-BY-SA license because ODbL does not grant a license to use the individual map in the database.[143]

In addition, these OGD licenses are mostly devised with a compatibility provision to make the subject license compatible with other similar public licenses.[144] Compatibility, or interoperability, between licenses means users can legally combine works subject to different public licenses together.[145] License compatibility is especially important in scientific fields, such as environmental protection and climate change, where users have an urgent need to use data from sources with different

---

or intent of restricting the ability of any person to exercise those rights. b. Parallel distribution. You may impose terms or technological measures on the Database, a Derivative Database, or the whole or a Substantial part of the Contents (a "Restricted Database") in contravention of Section 4.74 a. only if You also make a copy of the Database or a Derivative Database available to the recipient of the Restricted Database: i. That is available without additional fee; ii. That is available in a medium that does not alter or restrict the terms of this License, or any rights granted under it, or have the effect or intent of restricting the ability of any person to exercise those rights (an "Unrestricted Database"); and iii. The Unrestricted Database is at least as accessible to the recipient as a practical matter as the Restricted Database.").

140. Open Data Commons, ODbL, *supra* note 118, § 2.4 ("The individual items of the Contents contained in this Database may be covered by other rights, including copyright, patent, data protection, privacy, or personality rights, and this License does not cover any rights (other than Database Rights or in contract) in individual Contents contained in the Database. For example, if used on a Database of images (the Contents), this License would not apply to copyright over individual images, which could have their own separate licenses, or one single license covering all of the rights over the images.").

141. *See generally* Mordechai (Muki) Haklay & Patrick Weber, *OpenstreetMap: User-Generated street Maps*, 7(4) IEEE PERVASIVE COMPUTING 12 (2008); Pascal Neis & Alexander Zipf, *Analyzing the Contributor Activity of a Volunteered Geographic Information Project — The Case of OpenStreetMap*, 2012(1) ISPRS INT'L J. GEO-INFO. 146 (2012).

142. *See, e.g.*, Pascal Neis & Dennis Zielstra, *Recent Developments and Future Trends in Volunteered Geographic Information Research: The Case of OpenStreetMap*, 6(1) FUTURE INTERNET 76, 79 (2014).

143. Guibault & Margoni, *supra* note 114, at 158.

144. *See, e.g.*, Open Data Commons, *ODbL*, *supra* note 118, § 4.4(e); Mewhort, *supra* note 6, at 3, 20–21 (noting that the British Open Government Licence was intentionally crafted to be compatible with CC licenses).

145. *See, e.g.*, Brown, *supra* note 103, at 772–74; Lee, *supra* note 90, at 32–33; Martynas Mockus & Monica Palmirani, *Open Government Data Licensing Framework*, in ELECTRONIC GOVERNMENT AND THE INFORMATION SYSTEMS PERSPECTIVE: 4TH INTERNATIONAL CONFERENCE, EGOVIS 2015 287, 290–92 (Andrea Kő & Enrico Francesconi eds., 2015); Morando, *supra* note 85, at 445–48.

licenses.[146]  Because public licenses and declarations aim to facilitate greater distribution and reuse of the subject materials, the public licenses introduced above all include a compatibility provision so that users can legally combine content licensed under different licenses.[147]

To continue our analysis of various licensing terms, we focus below on issues of charges, restrictions on data usage, and the waiver of moral right, which open licenses approach differently.

1. Charges

Public licenses, such as those provided by CC and Open Data Commons, typically promote free sharing and use of materials.[148] Therefore, users or licensees do not need to pay for the licensed materials. Free of charge is normally a general principle found in OGD policies. For example, the New Zealand government made it clear in its NZGOAL that the "[c]harging by State Services agencies for people's use and re-use of copyrighted works and non-copyright materials is generally discouraged."[149]  Among all the OGD licenses and statements mentioned above, only the UK's Charged Licence charges users to use government data. A reasonable explanation for such a difference is that the Charged Licence was designed by the government and the government has certain practical considerations to reflect on, including the cost of implementing open data policies. It should also be noted that the UK government has deliberately placed two restrictions on the adoption of the Charged Licence: (1) this license is an exception; and (2) charges should be limited to the costs arising from "the re-use of information."[150]

Charging a reasonable fee for the use of government data is also permitted in the EU PSI Directive. According to the Directive, the fee is limited to "the marginal costs incurred for their reproduction, provision and dissemination," and the charges "shall not exceed the cost of collection, production, reproduction and dissemination, together with a reasonable return on investment."[151]  Although both the Charged Licence and the EU PSI Directive allow charging for the use of government data, the EU PSI Directive conflicts with the public interest concerns of OGD policy.  The Directive permits using open data as a tool to collect "a reasonable return on investment" other than "the cost of collection, production, reproduction and dissemination."[152] However, since the policy's goal is to promote transparency, accountability, participation, and economic development,[153] open data should not be used as a finance tool to benefit the government.[154] Therefore, the charges provision in the PSI Directive is obviously not the best practice for OGD policy.

---

146. *See, e.g.*, Estelle Derclaye, *The Role of Copyright in the Protection of Environment and the Fight Against Climate Change: Is the Current Copyright System Adequate?*, 5(2) WIPO J. 152, 156–58 (2014).
147. *See, e.g.*, Nat'l Archives, *OGL Version 3*, *supra* note 136.
148. *See, e.g.*, Armstrong, *supra* note 90, at 365–68; Pamela Samuelson, *Enriching Discourse on Public Domains*, 55 DUKE L.J. 783, 800–01 (2006).
149. NEW ZEALAND GOVERNMENT, *supra* note 102, at 16.
150. *See supra* text accompanying note 131.
151. EU Amending Directive on Re-Use of PSI, *supra* note 33, art. 6., at 12.

2. Restrictions on the Use of Data

In order to maximize the use of government data, a substantial segment of the open data community suggests that licensing terms should be the least restrictive or subject to minimal constraints.[155] Nevertheless, "minimal constraint" does not mean no constraints at all.[156] Accordingly, it becomes an issue as to what constitutes "minimal constraint" when the policy goal is to maximize the use of government data. Attribution is the most common restriction in public licenses.[157] Other than attribution, the Non-Commercial Government Licence under the UKGLF prohibits licensees from using or reusing the data for commercial purposes.[158] Since fostering innovation, new business models, and economic development are some of the primary policy goals of OGD, such licenses as the Non-Commercial Government Licence must be viewed as an exception. Otherwise, it would be impossible to launch new services or products based on government data at all.[159] Similar criticisms have been levied against the Italian Open Data License for its exclusion of commercial use.[160]

Moreover, the CC-BY-SA license and ODbL contain share-alike provisions, requiring the licensee to apply the same license to the adapted database or works created, as used in the original source.[161] Such provisions originate from the GNU General Public License (GPL) used in free software communities, where they hope to prevent licensees from hiding the modified code to gain unfair advantages.[162] This requirement prevents licensees from not giving back to the commons.[163] Share-alike provisions thus serve the function of broadening the commons for public use.[164] Nonetheless, governments have different considerations when licensing their data. A share-alike provision may impede new business models and innovative commercial uses of government data, which will eventually run counter to the policy goal of promoting economic development.[165] In other words, a share-alike duty may create unnecessary costs for enterprises that endeavor to develop novel products or services. Moreover, it is the responsibility of governments, rather than the private sector, to keep government data freely available. Even if the licensee does not honor the share-alike obligation, the same government data, dataset, or database is still open to the public. Therefore, CC-BY-SA licenses and ODbL are probably not the best options for a wide range of government data if economic growth is to remain a primary policy goal for OGD.

---

152. *Id.*
153. *See supra* Part II.
154. *See, e.g.*, Chris Corbin, *PSI Policy Principles: European Best Practice*, *in* ACCESS TO PUBLIC SECTOR INFORMATION (VOLUME 1) 161, 167 (2010), *available at* ses.library.usyd.edu.au// bitstream/2123/6561/1/PSI_vol1_chapter8.pdf.
155. *See supra* text accompanying note 33.
156. *But see* Chui et al., at 164 (claiming unrestrictive rights to use government data).
157. *See infra* Section IV.C.
158. *See supra* text accompanying note 130.
159. *See also* LEE ET AL., *supra* note 4, at 66 (suggesting that noncommercial provision is not acceptable in OGD licenses).
160. *See, e.g.*, Janssen, *supra* note 1, at 452.
161. *See supra* text accompanying notes 119, 137.
162. *See, e.g.*, Note, *On Enforcing Viral Terms*, 122 HARV. L. REV. 2184, 2187 (2009).
163. *See, e.g.*, Asay, *supra* note 91, at 760.

In summary, a government's choice of open licensing terms is quite different from that of the private sector. Businesses or communities usually link the choice over terms to contributors' incentives to contribute, costs to provide this incentive, and the sustainability of the commons' projects.[166] However, such considerations may not exist in the context of government data that is continuously generated as a government functions. In other words, a government's selection of open data licenses needs to reflect its policy goals, which are typically not addressed in private business or commons settings.

3. Waiver of Moral Rights

CC0 and PDDL are both waivers of copyright to the public domain. Surrender of copyright concerns moral rights issues in many countries. Although common law countries allow the waiver of moral rights,[167] those rights are not waivable in some civil law jurisdictions, like France.[168] In

---

164. *Cf.* Guy Pessach, *Reciprocal Share-Alike Exemptions in Copyright Law*, 30 CARDOZO L. REV. 1245, 1257–58 (2008) (noting the benefit of legislating reciprocal ShareAlike exemptions in copyright law).

165. *But see* LEE ET AL., *supra* note 4, at 104 (stating that ShareAlike provisions are acceptable in OGD licenses).

166. *See, e.g.*, Asay, *supra* note 91, at 773–80; *see also* Jyh-An Lee, *Organizing the Unorganized: The Role of Nonprofit Organizations in the Commons Communities*, 50 JURIMETRICS J. 275, 313 (noting that commons communities can sustain by using licensing terms to coordinate individual contributors).

167. *See, e.g.*, Gerald Dworkin, *The Moral Right of the Author: Moral Rights and the Common Law Countries*, 19 COLUM.-VLA J.L. & ARTS 229, 244–45 (1995); Henry Hansmann & Marina Santilli, *Authors' and Artists' Moral Rights: A Comparative Legal and Economic Analysis*, 26 J. LEGAL STUD. 95, 124–25 (1997); Russ VerSteeg, *Federal Moral Rights for Visual Artists: Contract Theory and Analysis*, 67 WASH. L. REV. 827, 845 (1992).

168. *See, e.g.*, PAUL GOLDSTEIN & BERNT HUGENHOLTZ, INTERNATIONAL COPYRIGHT: PRINCIPLES, LAW AND PRACTICE 367–68 (3d ed. 2013); Adolf Dietz, *Moral Rights and the Civil Law Countries*, 19 COLUM.-VLA J.L. & ARTS 199, 220–21 (1995); Neil Netanel, *Alienability Restrictions and the Enhancement of Author Autonomy in United States and Continental Copyright Law*, 12 CARDOZO ARTS & ENT. L.J. 1, 48–49 (1994); *see also* Ilhyung Lee, *Toward An American Moral Rights* other civil law jurisdictions, like Germany, moral rights can be waived under some circumstances.[169] Both CC0 and PDDL recognize the potential moral rights waiver issues present in civil law jurisdictions. However, their approaches differ slightly. CC0 adopts a more ambiguous tone by not mentioning a moral rights waiver in either its waiver or fallback provisions:

> 2. Waiver. To the greatest extent permitted by, but not in contravention of, applicable law, Affirmer hereby overtly, fully, permanently, irrevocably and unconditionally waives, abandons, and surrenders all of Affirmer's Copyright and Related Rights and associated claims and causes of action . . . .
> 3. Public License Fallback. Should any part of the Waiver for any reason be judged legally invalid or ineffective under applicable law, then the Waiver shall be preserved to the maximum extent permitted taking into account Affirmer's express Statement of Purpose.[170]

The courts may interpret the above CC0 provisions to maintain the validity of the waiver of economic rights alone, but not moral rights. If the attempt to waive copyright fails, CC0 confers an unconditional license permitting free reuse of the works and a covenant not to sue.[171] PDDL, on the other hand, copes with moral rights issues in a more sophisticated and clear manner. Section 3.4.b of the PDDL states:

> If waiver of moral rights under Section 3.4 a in the relevant jurisdiction is not possible, Licensor agrees not to assert any moral rights over the Work and waives all claims in moral rights to the fullest extent possible by the law of the relevant jurisdiction under Section 6.4.[172]

If we compare CC0 and PDDL in terms of a moral rights waiver, PDDL may provide more scope for users. Although moral rights in some civil law jurisdictions are unwaivable and nontransferrable, the licensee can still enjoy a certain degree of freedom associated with moral rights subject to the license agreement.[173] Therefore, a license agreement may alleviate the absolute unwaivable and non-transferrable moral rights in civil law jurisdictions. From this perspective, PDDL is more sophisticated and flexible than CC0 when it comes to coping with moral rights issues.

---

*in Copyright*, 58 WASH. & LEE L. REV. 795, 803 (2001) ("[j]urisdictions with the most advanced moral rights protection provide that moral rights are inalienable … and thus not subject to the author's transfer or waiver"). *But see* SILKE VON LEWINSKI, INTERNATIONAL COPYRIGHT LAW AND POLICY 53 (2008) (noting that moral rights are usually waivable in countries with those statutory rights).

    169. *See, e.g.*, Dietz, *supra* note 168, at 220.

    170. *CC01.0 Universal*, CREATIVE COMMONS, § 2–3, *available at* creativecommons.org/publicdomain/zero/1.0/legalcode (last visited Mar. 15, 2017).

    171. *Id.* § 3.

    172. Open Data Commons, PDDL, *supra* note 121, § 3.4(b).

    173. *See, e.g.*, GOLDSTEIN & HUGENHOLTZ, *supra* note 168, at 369.

IV. Legal Issues Underlying Licensing Government Data

Licensing government data involves a number of fundamental IP issues. Some IP issues are common for the legal governance of data, datasets, and databases, while others are uniquely associated with the nature of governments or open data policies. Below, this part discusses major IP issues in the government licensing of open data, which include the IP status of government data, the public domain nature of most government data, and the common attribution and non-endorsement provisions in public licenses.

IP STATUS OF GOVERNMENT DATA

Some scholars and policymakers assert that from a policy perspective, the works created by state employees should be in the public domain.[174] For example, the Dutch Council of State once opined that the City of Amsterdam could not legally impose any restriction on a company's use of the City's database because it was built with tax money.[175] In other words, the City's government did not own the database. In countries like the US, there are statutory public domain rules that prohibit the federal government from copyrighting works it produces; however, governments may still own copyrights assigned by others.[176] Even if a government can own a copyright, the originality standard may prevent it from owning a copyright over government data.[177] Originality is the universal standard for copyright protection.[178] A work needs to contain a minimum degree of creative authorship to be copyrightable.[179] Facts, or information automatically generated by a machine or algorithm, cannot be protected under copyright because they lack originality.[180] Database creators gain copyright protection of compilations and databases only if the selection, coordination, or arrangement of the contents is sufficiently original.[181] However, this is not the case for most government data or databases which include statistics, census data, fiscal data, budget information, parliamentary records, election records, property registration, facts about school locations and performance, other factual information, or are created automatically by machine.[182] In other words, government data usually lacks originality and therefore cannot be protected by copyright.[183]

---

174. Okediji, *supra* note 86, at 338–39; *see also* GURIN, *supra* note 4, at 9 ("governments should make the data they collect available to taxpayers who've paid to collect it"); Marcowitz-Bitton, *supra* note 2, at 415 (introducing the argument that government works "should be accessible to all, uninhibited by the restrains of copyright law, because the public sponsors the creation of these works with its tax money"); Ubaldi, *supra* note 1, at 40 (noting the reason of this debate is that government "information has been created with tax-payers' money"). *But see* Khayyat & Bannister, *supra* note 20, at 244 (rebutting the argument that taxpayers shall have free access to government generated data).

175. Janssen, *supra* note 1, at 451.

176. 17 U.S.C. § 105 (2012) ("[c]opyright . . . is not available for any work of the United States Government"); *see also* Marcowitz-Bitton, *supra* note 2, at 420 (explaining that "[t]he reason behind 17 U.S.C. § 105 is to ensure that government information remains in the public domain in order to best serve the public interest); Okediji, *supra* note 86, 343–45 (explaining the evolution of public domain rule on government works in the U.S.).

177. *See, e.g.*, Beth Ford, Comment, *Open Wide the Gates of Legal Access*, 93 OR. L. REV. 539, 546 (2014).

178. *See, e.g.*, GOLDSTEIN & HUGENHOLTZ, *supra* note 168, at 194–93.

179. *See, e.g.*, WILLIAM CORNISH ET AL., INTELLECTUAL PROPERTY: PATENTS, COPYRIGHT, TRADE MARKS AND ALLIED RIGHTS 435 (8th ed. 2013); *see also* JULIE COHEN ET AL., COPYRIGHT IN A GLOBAL INFORMATION ECONOMY 57 (3d ed. 2010) ("nearly all countries require some level of creativity as a prerequisite for copyright protection").

In the EU, databases can be protected by a *sui generis* right, which is independent from copyright under the European Database Protection Directive.[184] In order to obtain the *sui generis* protection, the database owner needs to prove that a substantial investment was made in the database.[185] The right is granted for fifteen years and allows the owner to control who can use the database, how the database can be used, and the database's distribution.[186] Therefore, it is easier for EU governments (or those with similar laws) to own their databases' IP rights and therefore maintain more legal control over their data. Put differently, European governments can still use public licenses, such as CC-BY, ODC-BY, or ODbL, to legally license their database right even if their data is not copyrightable.

---

180. *See, e.g.*, FREDERICK M. ABBOTT ET AL., INTERNATIONAL INTELLECTUAL PROPERTY IN AN INTEGRATED WORLD ECONOMY 535 (3rd ed. 2015); Scassa, *supra* note 4, at 1782–83, 1787–88.
181. *See. e.g.*, Feist Publications, Inc. v. Rural Tel. Serv. Co., 499 U.S. 340, 345 (1991).
182. *See. e.g.*, Marcowitz-Bitton, *supra* note 2, at 415; Okediji, *supra* note 86, at 334; *see also* TAUBERER, *supra* note 1, at 115 ("government data normally represents facts about the real world (who voted on what, environmental conditions, financial holdings)"); Borgesius et al., *supra* note 17, at 2094 (noting that government data includes statistics, land registries, business registers, or earth observation data); Marcowitz-Bitton, *supra* note 2, at 413 (stating that government data includes "national statistics, budget information, parliamentary records, data about the location of schools and their performance, information about crimes, election records, financial data, and more"); LEE ET AL., *supra* note 4, at 54–56 (noting that common high-value datasets are those of company register, insolvency and bankruptcy record, government contract, various statistics, and so on); Ubaldi, *supra* note 1, at 6, 23 (noting that government data consists of business information, registers, geographic information, meteorological information, social data on statistics, and transport information).
183. *See, e.g.*, Marcowitz-Bitton, *supra* note 2, at 415; Scassa, *supra* note 4, at 1785–86; *see also* Paul Miller et al., *Open Data Commons, A License for Open Data*, (369 CEUR WORKSHOP PROCEEDINGS, no. 8, Apr. 22, 2008), *available at* ceur-ws.org/Vol-369/paper08.pdf (similarly holding that data, datasets, and databases are mostly not copyrightable creative works); Scassa, *supra* note 4, at 1766 (stating that it is extremely difficult to identify authorship in government data).
184. Directive 96/9/EC of the European Parliament and of the Council of 11 March 1996 on the Legal Protection of Databases, 1996 O.J. (L 77) 20, 20, *available at* eur-lex.europa.eu/legal-content /EN/ TXT/?uri=CELEX%3A31996L0009.
185. *Id.* art. 7, at 25.
186. *Id.* art. 10, at 26.

Winter 2017]     *LICENSING OPEN GOVERNMENT DATA*     231Data and databases may be protected as trade secrets as well.[187] However, by its very nature, trade secret protection cannot be applied to OGD. Secrecy is the defining element of a trade secret.[188] Once information is known by the public, it is no longer secret and consequently cannot be protected as a trade secret.[189] Since transparency is one of the policy goals of OGD,[190] such data needs to be openly accessible.[191] The openness of government data certainly fails to fulfill the secrecy requirement in trade secret law.[192] Therefore, it is impossible to protect OGD as trade secrets.

Data holders or database owners occasionally use contract and/or technical restrictions to control access to their data or databases.[193] It does not matter if the data is in the public domain because those database owners do not claim IP over the data or databases. The Lexis and Westlaw databases are good examples of databases with restricted access to public information. Both contain huge amounts of data related to court decisions and other legal texts that are in the public domain, but only subscribers have access to these databases.[194] However, governments cannot use the same approach to control open data. In the OGD context, digital technologies are used to disseminate government data,[195] rather than restrict access to it. Moreover, the closed nature of the Lexis and Westlaw databases enables better technological and contractual control. Subscribers

---

187. *See, e.g.*, Jennifer Askanazi et al., *The Future of Database Protection in U.S. Copyright Law*, 2001 DUKE L. & TECH. REV. 17, ¶15 (2001); Donna M. Gitter, *Resolving The Open Source Paradox in Biotechnology: A Proposal for A Revised Open Source Policy for Publicly Funded Genomic Databases*, 43 HOUS. L. REV. 1475, 1513 (2007); Lionel M. Lavenue, *Database Rights and Technical Data Rights: The Expansion of Intellectual Property for the Protection of Databases*, 38 SANTA CLARA L. REV. 1, 26 (1997); Angela M. Oliver, *Personalized Medicine in the Information Age: Myriad's De Facto Monopoly on Breast Cancer Research*, 68 SMU L. REV. 537, 648–50 (2015).

Sharon K. Sandeen, *A Contract by Any Other Name Is Still A Contract: Examining the Effectiveness of Trade Secret Clauses to Protect Databases*, 45 IDEA 119, 162–63 (2005).

188. Jonathan R. Chally, *The Law of Trade Secrets: Toward A More Efficient Approach*, 57 VAND. L. REV. 1269, 1283-84 (2004); Gitter, *supra* note 187, at 1510; *see also* Mark Lemley, *The Surprising Virtues of Treating Trade Secrets as IP Rights*, 61 STAN. L. REV. 311, 342–43 (2008) (illustrating the centrality of secrecy in trade secret law).

189. Lavenue, *supra* note 187, at 3; Sandeen, *at* 133–34.

190. *See* text accompanying note 61-71.

191. *See* text accompanying note 32-47.

192. *Cf* David S. Levine, *Secrecy and Unaccountability: Trade Secrets in Our Public Infrastructure*, 59 FLA. L. REV. 135, (2007) (arguing that "trade secrecy must give way to traditional notions of transparency and accountability when it comes to the provision of public infrastructure").

193. Pamela Samuelson, *Mapping the Digital Public Domain: Threats and Opportunities*, 66 LAW & CONTEMP. PROBS. 147, 152 (2003); *see also* J. H. Reichman & Paul F. Uhlir, *A Contractually Reconstructed Research Commons for Scientific Data in a Highly Protectionist Intellectual Property Environment*, 66 LAW & CONTEMP. PROBS. 315, 401 (2003) ("the data that traditional copyright law puts into the public domain may be fenced to a still unknown extent by the technological measures").

194. Samuelson, *supra* note 193, at 152; Jason Mazzone, *Copyfraud*, 81 N.Y.U. L. REV. 1026, 1046 (2006).

195. *See supra* text accompanying note 5.

to the Lexis and Westlaw databases are limited and identifiable, whereas users of OGD are not. As a result, it is impractical for governments to govern the use of open data via contractual and technical restrictions.

In sum, government data is not protected by copyright unless it meets the originality standard in copyright law. Although some government data is an original expression subject to copyright protection, most government data is not, such as statistics, factual information, or information automatically produced by machine or algorithm. Governments in EU countries can obtain *sui generis* protection for their databases if they made a substantial investment into the creation of the databases. However, this *sui generis* right is not available in most other jurisdictions. As a result, licenses that are specifically designed to cover a database right, such as ODbL, are more suited to EU countries than others.[196]

LEGAL EFFECT OF LICENSING DATA IN THE PUBLIC DOMAIN

Typically, works subject to open licenses are protected by copyright or other types of IP rights. Accordingly, most open data licenses are designed based on the presumption that the subject government's data is copyrighted.[197] For example, the UKGLF makes it clear that it only applies to copyright and database rights.[198] It implies that the UKGLF will not be applied to data that is not protected by copyright or database rights. On the flip side, materials in the public domain do not require licenses to be released.[199] Nonetheless, what if governments apply public licenses to data that is not protected by copyright or database rights? What would be the legal effect of such licenses?

Although public licenses, such as CC licenses, can help disseminate copyrighted work legally, they are not required to release data that is already in the public domain.[200] Some researchers have rightfully pointed out that CC licenses are not tailor-made for non-copyright materials.[201] Instead, governments may use a PDM to clarify the public domain status of the data. If governments use CC0 for such data, it may not cause any serious legal problem, although the underlying data is not copyrighted at all. In this regard, CC0 has the advantage of providing users with more legal certainty because it can show the public that the affirmer is committed to relinquishing their IP protection, if they have it, to the broadest possible extent.[202] Users mostly do not need to seek legal advice regarding data released under CC0.[203] A number of US government agencies have used CC0 to release data or put it in the public domain.[204]

---

196. *See, e.g.*, Mewhort, *supra* note 6, at 3 (noting that the UK Open Government Licence is more appropriate for EU countries where *sui generis* database rights exists).
197. Scassa, *supra* note 4, at 1804; *see also* TAUBERER, *supra* note 1, at 107 ("[w]hen a work is copyrighted, a license is required to undo or partially undo the all-rights-reserved default rule"), and at 144 ("[o]pen licensing…is subject to copyright protections"); Bannister, *supra* note 61, at 1099 ("Creative Commons licensing movement aims to provide a standardised infrastructure for the open licensing of copyright protected material").
198. NAT'L ARCHIVES, UK GOVERNMENT LICENSING FRAMEWORK, *supra* note 1, at 6.
199. *See, e.g.*, Derclaye, *supra* note 146, at 156 ("[i]n countries where official texts are not protected by copyright, the issue of the need for access through licenses does not even arise").
200. TAUBERER, *supra* note 1, at 107.
201. Korn & Oppenheim, *supra* note 89, at 4, 6.

Yet, applying public licenses, such as CC, to non-copyrightable or public-domain data is not an uncommon OGD practice. For example, the Bureau of Meteorology in Australia releases weather observation datasets with a CC-BY 3.0 Australia license.[205] This data is updated every 30 minutes.[206] However, this dataset primarily provides factual information, such as temperature and humidity,[207] which is not copyrightable. Similarly, the New Zealand government has released public holidays and anniversary dates, which are non-copyrightable factual information, under a CC-BY 3.0 license.[208] Consequently, these practices lead to a legal problem pertaining to how governments can use CC or other public licenses to release non-copyrightable data in countries where database rights are not protected. It will take further empirical study to explore why governments in non-EU jurisdictions tend to adopt public licenses, rather than CC0, PDM, or PDDL, to release public domain data. It is possible that governments do not conduct due diligence regarding the legal status of the subject data. Based on the author's personal experiences of providing OGD consultation to the public sector, it is more likely because of the governmental mentality regarding control over data. Government officials may hesitate to recognize the public domain nature of the sorts of data over which they used to exert their full control. They may not understand that although the government is in charge of data governance, the government cannot legally claim data ownership.

From a legal perspective, it is worthwhile to explore the effects of these open data licenses if the underlying data is in the public domain. There are two possible approaches to this question. The first interpretation is that the contract may become void or partly void given the subject matter data in the contract is not owned by the licensor and is in the public domain. The data should thus be free to everyone.[209] If the underlying data is not protected under applicable law, a license is needless.[210] As a result, asserting copyright over public domain materials may at worst be defined as "copyfraud," which may stifle creativity and free speech.[211]

---

202. *See supra* text accompanying note 105.
203. Mewhort, *supra* note 6, at 17.
204. *See, e.g.*, WHITE HOUSE, U.S. OPEN DATA ACTION PLAN (2014), *available at* obamawhitehouse.archives.gov/sites/default/files/microsites/ostp/us_open_data_action_plan.pdf.
205. Bureau of Meteorology (Australia), *Latest Coastal Weather Observations for Coolangatta (QLD)*, DATA.GOV AUSTRALIA, *available at* data.gov.au/dataset/latest-coastal-weather-observations-for-coolangatta-qld (last visited Mar. 15, 2017).
206. *Id.*
207. *See, e.g.*, Bureau of Meteorology (Australia), *Latest Weather Observations for Coolangatta*, DATA.GOV AUSTRALIA, *available at* bom.gov.au/products/IDQ60801/IDQ60801.94592.shtml (last visited Mar. 15, 2017).
208. Ministry of Business Innovation and Employment (New Zealand), *Public Holidays*, DATA.GOVT NEW ZEALAND (Aug. 26, 2016), *available at* data.govt.nz/dataset/show/5686 (using a CC "Attribution 3.0 New Zealand" license).

The second possible solution is to recognize the validity of the agreement and treat it as a binding contract between the data holder and the user. Put differently, even if there are no underlying IP rights to constitute a license,[212] the agreement itself is still a contract that can legally oblige data users to fulfill attribution, share-alike, or other duties.[213] This contract theory was criticized as imposing unnecessary restrictions on public domain resources.[214] Even if the agreement is a valid contract, government agencies may not be able to enforce it against users who breach the contract. Many open data agreements provide a provision that data users' rights will be revoked or the agreement will be terminated automatically if the users do not comply with the conditions.[215] This is typically the only legal effect of licensees' non-compliance. However, if this is the only legal effect of non-compliance, it will make no difference to noncompliant users. Those users can always argue that they have the intrinsic right to use data in the public domain even without an agreement in place.

Enforceability has long been an issue for public licenses, such as CC licenses.[216] Even if the open license agreement is valid and enforceable between the licensor and licensee, whether the data is protected by IP makes a big difference when the licensor enforces their legal right against a third party. Public licenses, like CC licenses, are merely contracts with less effect than the law.[217] Given the transparent nature of OGD policy, it is quite possible that third parties do not obtain the data directly from the government, but from elsewhere. These third parties may argue that they are not parties to public licenses and thus are not bound by the license agreement. Such risk is higher than that in a proprietary licensing scenario,

---

209. *See, e.g.*, Marcowitz-Bitton, *supra* note 2, at 438.
210. *Comments on the Open Database License Proposed by Open Data Commons*, CREATIVE COMMONS, *available at* sciencecommons.org/resources/readingroom/comments-on-odbl/ (last visited Mar. 15, 2017).
211. Mazzone, *supra* note 194, at 1028–30.
212. *Cf.* Christopher M. Newman, *A License Is Not a "Contract Not to Sue": Disentangling Property and Contract in the Law of Copyright Licenses*, IOWA L. REV. 1101, 1114 (2013) (noting that a license needs to be granted by the title holder of the property).
213. Creative Commons, *supra* note 210.
214. *Id.*
215. *See, e.g.*, Commission Notice, Guidelines on Recommended Standard Licences, *supra* note 93, § 2.3.6; *Attribution 4.0 International*, CREATIVE COMMONS, § 6, https://creativecommons.org/licenses/by/4.0/legalcode (last visited Mar. 15, 2017); Creative Commons, *Attribution-ShareAlike 4.0 International*, *supra* note 137, § 6; Open Data Commons, *ODC-By v1.0*, *supra* note 116, § 9.1.
216. *See, e.g.*, Brown, *supra* note 103, at 767.
217. *See, e.g.*, MERGES, *supra* note 90, at 229.

where the number of licensees is limited. Traditionally, even though IP owners cannot sue the third parties for breach of the license agreement, they can still claim IP infringement against them.[218] Nevertheless, if the subject data is in the public domain, the data owners certainly do not have any grounds to sue the third party who does not comply with the license agreement.

*Attribution*

Open licenses with attribution requirements alone, such as CC-BY, ODC-BY, and the Open Government Licence, are generally the most permissive licenses. Therefore, some commentators view these attribution-only licenses as "quasi-public domain dedications."[219] In the Guidelines on Recommended Standard Licences, the European Commission proposed two acceptable restrictions to open data licenses, which are "acknowledgment of source" and "acknowledgment of any modifications to the document."[220] The Guidelines further explained that "any other obligations [than attribution] may limit licensees' creativity or economic activity, thereby affecting the re-use potential of the documents in question."[221] In this section, the common attribution requirement in most OGD policies and licenses is explored and the theory of attribution in moral rights is used as a lens to understand the rationale behind attribution provisions in open data licenses.

*Proper Attribution*

Almost all public licenses or open licenses contain an attribution requirement.[222] CC0, PDM, and PDDL are probably the only three open data terms that do not require attribution. Some government agencies use CC0 to waive all their copyright and related rights, but still require users to give attribution. For example, the Terms of Service provided by the US

---

218. *See also* Brown, *supra* note 103, at 767 ("if the user [of a CC license] cannot rely on the license then they will have no way to know whether their use constitutes copyright infringement").
219. *See, e.g.*, Asay, *supra* note 91, at 760.
220. Commission Notice, Guidelines on Recommended Standard Licences, *supra* note 93, § 2; *see also* TAUBERER, *supra* note 1, at 109 (discussing the contractual obligation associated with "attribution and data integrity").
221. Commission Notice, Guidelines on Recommended Standard Licences, *supra* note 93, § 2.3.2.
222. *See e.g.* Catherine L. Fisk, *Credit Where It's Due: The Law and Norms of Attribution*, 95 GEO. L.J. 49, 90–92 (2006); Eric E. Johnson, *Rethinking Sharing Licenses for the Entertainment Media*, 26 CARDOZO ARTS & ENT. L.J. 391, 409 (2008); Greg Lastowka, *Digital Attribution: Copyright and the Right to Credit*, 87 B.U. L. REV. 41, 59, 78–84 (2007); *see also* Mira T. Sundara Rajan, *Creative Commons: America's Moral Rights?*, 21 FORDHAM INTELL. PROP. MEDIA & ENT. L.J. 905, 925 (2011) (noting that attribution is a fundamental condition for CC licenses).

Food and Drug Administration (FDA) makes CC0 the default rule for open data while requiring users give "proper credit."[223] The legal outcome of such an arrangement is similar to that of adopting CC-BY or ODC-BY, which grants virtually all types of copyright and related rights as long as the licensors attribute any public use of the database, or works produced from the database, in the manner specified in the license. But why does the FDA not apply CC-BY directly to ensure user attribution? This practice would involve a fundamental inquiry into the relationship between IP and attribution. Normally, right of attribution is part of moral rights. Authors own copyright so that they can require users or licensees to attribute credit to them. In the case of the FDA's Terms of Service, the agency does not intend to claim copyright while it values users' attributions. It may well explain why government is interested in claiming ownership over non-copyrightable materials.[224] By claiming copyright ownership over data, governments are justified in using public licenses to entail users giving credit.

Since OGD policies normally promote access to, and reuse of, data for free or at nominal costs,[225] every restriction in the licensing terms that increases users' costs needs to be justified. Therefore, it is worth exploring why attribution is necessary in open data licenses. Some researchers argue that the attribution requirement is the government's instrument to control speech because every restriction on the use of data is a form of censorship.[226] This argument is flawed in at least three ways: first, free speech as a constitutional right is still subject to some limitations;[227] second, there is no empirical evidence or theoretical support indicating that the attribution requirement in OGD licenses generates a chilling effect or any barriers to freedom of speech; and third, it is not articulated why governments would intend to restrict speech via the attribution requirement. We can hardly imagine how a government would be able to use the attribution requirement to silence others from voicing opinions with which it disagrees.

Some other scholars have suggested that attribution can guarantee the accuracy and reliability of the data provided by governments.[228] Nevertheless, such an argument may not be validated if we read through government data licenses. It is quite costly to maintain the accuracy and precision of data.[229] Poor quality has been a problem for government data;[230] consequently, making it openly available highlights its incompleteness and inaccuracy. Most open data licenses include a liability disclaimer refusing to take responsibility for the data's accuracy, correctness, or completeness.[231] The data or database is licensed by the licensor "as is" and without any warranty of data quality.[232] The disclaimer provision in traditional public license agreements is typically subject to IP infringement claimed by third parties,[233] but in open data agreements, the disclaimer provision also excludes any legal liability associated with data error. If the attribution terms in open data licenses are intended to ensure data quality, then the disclaimer provisions become unnecessary in the license agreement.

---

223. FDA, *Terms of Services*, *supra* note 88.
224. *See supra* text accompanying note 205-208.
225. *See supra* text accompanying note 32.
226. TAUBERER, *supra* note 1, at 109.
227. *See, e.g.*, David S. Bogen, *The Origins of Freedom of Speech and Press*, 42 MD. L. REV. 429, 431, 436–37 (1983); Irene M. Ten Cate, *Speech, Truth, and Freedom: An Examination of John Stuart Mill's and Justice Oliver Wendell Holmes's Free Speech Defenses*, 22 YALE J.L. & HUMAN. 35, 69 (2010); Ronald J. Krotoszynski, Jr, *A Comparative Perspective on the First Amendment: Free Speech, Militant Democracy, and the Primacy of Dignity as a Preferred Constitutional Value in Germany*, 78 TUL. L. REV. 1549, 1551, 1554–59 (2004).
228. *See, e.g.*, Marcowitz-Bitton, *supra* note 2, at 414–15.

*Over-Attribution*

Although governments want proper attribution for the data they release, they dislike over-attribution. One notable example is the UK's OGL which encompasses a "non-endorsement" provision that prohibits the use of the released information "in a way that suggests any official status or that the Information Provider and/or Licensor endorse . . . [the licensor's] use of the Information." Similar provisions can also be found in the Dutch OGD policy[234] and in all CC licenses.[235] A nonendorsement provision addresses a typical concern that the public sector has with open data. When releasing data to the public, government agencies normally aim to release it in a non-discriminatory and neutral way. After all, encouraging innovative uses of the data by the private sector does not mean endorsing or recommending those uses.

---

229. TAUBERER, *supra* note 1, at 118.
230. *See, e.g.*, GURIN, *supra* note 4, at 233; TAUBERER, *supra* note 1, at 149.
231. *See, e.g.*, Commission Notice, Guidelines on Recommended Standard Licences, *supra* note 93, § 2.3.5; FDA, Terms of Services, *supra* note 88; Nat'l Archives, *OGL Version 3*, *supra* note 136; Open Data Commons, *ODC-By v1.0*, *supra* note 116, § 7.0; Open Data Commons, *PDDL*, *supra* note 121, § 5.0.
232. *See, e.g.*, LEE ET AL., *supra* note 4, at 69.
233. *See, e.g.*, Stephen McJohn, *The GPL Meets the UCC: Does Free Software Come with a Warranty of No Infringement?*, 15 J. HIGH TECH. L. 1, 19 (2014).
234. *See, e.g.*, Janssen, *supra* note 1, at 451.
235. Creative Commons, *supra* note 96.

238 HASTINGS BUSINESS LAW JOURNAL [Vol. 13:2

*Rationale for Attribution in Open Data*

The right to be identified, or right of attribution or paternity, is the most important category of moral rights.[236] Therefore, a government's attitude toward appropriate attribution may be understood from the perspective of moral rights theory. It should be noted that the government's generation of data differs from that of individuals or enterprises making creative works. Most government data is produced as a by-product of its daily functions.[237] Therefore, although correct attribution can provide non-pecuniary rewards or incentives to authors of creative works,[238] the same cannot be justified in the context of government data. In addition, attribution rights have traditionally represented an artist's personal connection to his or her creative works.[239] This personal link hardly exists in the generation of government data.

Nonetheless, governments occasionally gain political advantages from the attribution requirement because it helps craft the public impression that they have released some valuable data to society. In this sense, governments, just like authors of creative works, benefit from situations where the relationship between the makers and their works is visible.[240] Greg Lastowka correctly indicated that attribution helps creators gain advantages in the reputation market.[241] The same reasoning can be applied to governments' open data licenses in which the attribution requirement may help them earn a positive public reputation.

Another argument in favor of attribution is the "public interest theory" that states that the public can benefit from the disclosure of attribution.[242]

---

236. *See, e.g.*, GOLDSTEIN & HUGENHOLTZ, *supra* note 168, at 361; Rajan, *supra* note 222, at 926.

237. *See supra* text accompanying note 2.

238. *See, e.g.*, Jane C. Ginsburg, *Moral Rights in a Common Law System*, 1(4) ENT. L. REV. 121, 122 (1990); Fisk, *supra* note 222, at 56–60; *see also* Asay, *supra* note 91, at 792 (noting that attribution is a significant drive for contributions in free or open source software or open content communities).

239. *See, e.g.*, Elizabeth M. Bock, *Note: Using Public Disclosure as the Vesting Point for Moral Rights under the Visual Artists Rights Act*, 110 MICH. L. REV. 153, 161–62 (2011); *see also* Robert C. Bird, *Moral Rights: Diagnosis and Rehabilitation*, 46 AM. BUS. L.J. 407, 426 (2009) (*Le droit moral* [or moral right in France] . . . addresses legal rights that acknowledge a personal legal connection between an author and her creations); Dietz, *supra* note 168, at 207 (noting that morals rights in Germany Copyright Act focuses on the authors' personal relationship with his or her creative works).

240. *See* VON LEWINSKI, *supra* note 168, at 51.

241. Lastowka, *supra* note 222, at 60–61; *see also* Bock, *supra* note 239, at 168 ("integrity and attribution are concerned with the reputation of the artist").

242. *See, e.g.*, Ginsburg, *supra* note 238, at 122; *see also* Fisk, *supra* note 222, at 54 ("[a]ttribution is a type of signal, and it operates in labor and other markets plagued by information asymmetries in which reliable signals are important"); Hansmann & Santilli, *supra* note 167, at 107 (noting that public interests are enhanced by attribution right, which prevents the public from being misled about the work); Margaret Ann Wilkinson, *The Public Interest in Moral Rights Protection*, 2006 MICH. ST. L. REV. 193, 212–16 (2006) (analyzing moral rights' public-interests function in information provision).

This theory is more suited to OGD policy. As the primary goal of OGD is to promote transparency, accountability, and economic development, the public has a stake in knowing whether the data is provided by the government and which government agency provided which data, dataset, or database. The disclosure of this information can better enable citizens to assess the performance of government agencies and whether, and to what extent, the data release can help economic development.

The "public interest theory" may also justify the nonendorsement provision mentioned above. In many jurisdictions, moral rights are associated with not only a user's obligation to identify the author but also a nonauthor's right to object to false attribution.[243] Creators can prevent works that they never created from being misattributed to them.[244] This right is the reverse of attribution rights.[245] The right to object to false attribution is different from a nonendorsement scenario because the former did not create the subject information at all, whereas governments did create the data in the latter but they refuse to endorse private parties' uses of it. Nevertheless, users in both scenarios attempt to ride on the coattails of another's reputation and mislead the public in order to market their products or services. In this sense, the "public interest theory" can also explain the rationale for the non-endorsement provision in government licenses. Like the laws that prohibit false attribution, the nonendorsement provision helps prevent public deception.[246] It is in the public interest to prevent governments or any other parties from receiving undue attribution or false association.[247]

## V. Conclusion

Governments collect and generate a great deal of data as a part of their daily functions, and this data has tremendous public and private value. By enabling governments to release vast amounts of data in a timely manner, digital technologies propel the OGD movement. Open data may contribute to the achievement of a wide range of social, economic, and political goals. Nevertheless, it also involves a variety of legal issues. The choice, or design, of licenses for OGD is not only a legal issue, but also a policy issue. OGD licenses can form an important element of a government's information policy, reflecting considerations that differ from those of proprietary licenses or community-based commons licenses in the private sector. Therefore, this study argues that a government's decisions regarding open data licenses reveal the priorities of its policy goals, which may be associated with transparency, accountability, collaboration, or economic growth.

---

243. *See, e.g.*, CORNISH ET AL., *supra* note 179, at 514; Bird, *supra* note 239, at 411–12; Dworkin, *supra* note 167, at 232; Hansmann & Santilli, *supra* note 167, at 130; Lee, *supra* note 168, at 802; Cyrill P. Rigamonti, *Deconstructing Moral Rights*, 47 HARV. INT'L L.J. 353, 361, 401 (2006).

244. *See, e.g.*, TANYA APLIN & JENNIFER DAVIS, INTELLECTUAL PROPERTY LAW: TEXT, CASES, AND MATERIALS 151 (2nd ed. 2009); CORNISH ET AL., *supra* note 179, at 514; Robert C. Bird, Lucille M. Ponte *Protecting Moral Rights In the United States and the United Kingdom: Challenges and Opportunities under the U.K.'s New Performances Regulations*, 24 B.U. INT'L L.J. 213, 221, 236–38 (2006).

245. *See, e.g.*, Ginsburg, *supra* note 238, at 122.

246. *Id.*

247. *Id.*

In this Article, three suites of public licenses developed by CC, ODC, and the British government are compared and analyzed. They are probably the most notable licenses for government data and commonly considered by the public sector for open data policies. As a huge amount of government data does not meet the originality requirement and thus is not copyrightable, these licenses may not constitute effective copyright protections in many OGD scenarios. However, they can still function well as licenses of database rights in EU jurisdictions where database rights are protected as *sui generis* rights. In most other jurisdictions, such as Asian countries and the US, where there is no *sui generis* database right, these licenses may not be legally effective for non-copyrightable data and databases. In these cases, governments are advised to implement CC0, PDM, or PDDM after conducting due diligence confirming the public domain status of the subject data.

Moreover, attribution is the most common, and occasionally the only, requirement in OGD licenses. This requirement is typically accompanied by a nonendorsement provision. The existence of this design in open data licenses cannot be explained by traditional copyright theories because the data can hardly present a government's personality and governments do not need user attribution as an incentive to generate data. Nonetheless, these provisions can be understood by applying the public interest theory of moral right. As the primary goal of OGD is to promote transparency, accountability, and economic development, the public has a vested interest in knowing whether the data is provided by the government and which government agency has produced it. Moreover, governments need user attribution to earn a positive public reputation. The nonendorsement provision is also used to protect the public interest by preventing governments or any other parties from benefiting from undue attribution or false association.